\documentclass[preprint,showpacs,preprintnumbers,amsmath,amssymb]{revtex4}
\usepackage{epsfig}
\usepackage{graphicx}% Include figure files
\usepackage{dcolumn}% Align table columns on decimal point
\usepackage{bm}% bold math
\usepackage{threeparttable}
\usepackage{centernot}

\def\beq{\begin{equation}}
\def\eeq{\end{equation}}
\def\eeqn{\end{equation}}
\newcommand\iden{\leavevmode\hbox{\small1\normalsize\kern-.33em1}}

\newcommand{\bea} {\begin{eqnarray}}
\newcommand{\eea} {\end{eqnarray}}

\def\lam{\lambda}

\def\hm{{\hat m}}

\newcommand{\nn}{\nonumber}

\let\jnfont=\rm
\def\NPB#1 {{\jnfont Nucl.\ Phys.\ B }{\bf #1} }
\def\PLB#1 {{\jnfont Phys.\ Lett.\ B }{\bf #1} }
\def\EPJC#1 {{\jnfont Eur.\ Phys.\ Jour.\ C }{\bf #1} }
\def\PRD#1 {{\jnfont Phys.\ Rev.\ D }{\bf #1} }
\def\PRL#1 {{\jnfont Phys.\ Rev.\ Lett.\ }{\bf #1} }
\def\MPLA#1 {{\jnfont Mod.\ Phys.\ Lett.\ A }{\bf #1} }
\def\JPG#1 {{\jnfont J.\ Phys.\ G }{\bf #1} }
\def\CTP#1 {{\jnfont Commun.\ Theor.\ Phys.\ }{\bf #1} }
\def\JHEP#1 {{\jnfont JHEP \ }{\bf #1} }
\def\NPPS#1 {{\jnfont Nucl.\ Phys.\ Proc.\ Suppl.\ }{\bf #1} }
\def\CPC#1 {{\jnfont Comput.\ Phys.\ Commun.\ }{\bf #1} }
\def\CPL#1 {{\jnfont Chin.\ Phys.\ Lett. }{\bf #1} }
\def\APPB#1 {{\jnfont Acta\ Phys.\ Polon.\ B }{\bf #1} }

\def\lsim{\raise0.3ex\hbox{$<$\kern-0.75em\raise-1.1ex\hbox{$\sim$}}}
\def\gsim{\raise0.3ex\hbox{$>$\kern-0.75em\raise-1.1ex\hbox{$\sim$}}}
\def\PR#1 {{\jnfont Phys.\ Rept. }{\bf #1} }
\def\CHC#1 {{\jnfont Chin.\ Phys.\ C }{\bf #1} }
\def\NIMA#1 {{\jnfont Nucl.\ Instrum.\ Meth.\ A }{\bf #1} }
\def\JCAP#1 {{\jnfont JCAP \ }{\bf #1} }
\def\ASA#1 {{\jnfont Astron.\ Astrophys.\ A }{\bf #1} }  
\begin{document}

\title{\ \\[10mm] The $U(1)_{L_\mu-L_\tau}$ breaking phase transition, muon $g-2$,  dark matter, collider and gravitational wave}
,
\author{Jie Wang, Jinghong Ma, Jing Gao, Xiao-Fang Han$^{*}$\footnotetext{*) 
Corresponding author. Email address: xfhan@ytu.edu.cn}, Lei Wang}

\affiliation{Department of Physics, Yantai University, Yantai 264005, P. R. China}

%---------------------------------------------------------------------------

\begin{abstract}
Combining the dark matter and muon $g-2$ anomaly, we study the $U(1)_{L_\mu-L_\tau}$ breaking phase transition, gravitational wave spectra, and the direct detection at the LHC
in an extra $U(1)_{L_\mu-L_\tau}$ gauge symmetry extension of the standard model.
The new fields includes vector-like leptons ($E_1,~ E_2,~ N$), $U(1)_{L_\mu-L_\tau}$ breaking scalar $S$ and gauge boson $Z'$, as well as the dark
matter candidate $X_I$ and its heavy partner $X_R$. 
A joint explanation of the dark matter relic density and muon $g-2$ anomaly excludes the region where both $min(m_{E_1},m_{E_2},m_N,m_{X_R})$ and $min(m_{Z'},m_S)$ 
are much larger than $m_{X_I}$. 
In the parameter space accommodating the DM relic density and muon $g-2$ anomaly, the model can achieve a first-order $U(1)_{L_\mu-L_\tau}$ breaking phase transition, 
whose strength is sensitive to the parameters of Higgs potential.
The corresponding gravitational wave spectra can reach the sensitivity of U-DECIGO.
In addition, the direct searches at the LHC impose stringent bound on the mass spectra of the vector-like leptons and dark matter. 
\end{abstract}

\maketitle
\section{introduction}
An extra $U(1)_{L_\mu-L_\tau}$ gauge symmetry extension of the standard model (SM) is anomaly-free and
naturally breaks Lepton Flavour Universality (LFU) because the
$U(1)_{L_\mu-L_\tau}$ gauge boson couples only to  $\mu(\tau)$ but
not to e. The model was originally formulated by He, Joshi, Lew, and Volkas \cite{lu-lt}.
Thereafter, this type of $U(1)_{L_\mu-L_\tau}$ model has  been
modified from its minimal version, and many variants have been proposed in the context of different phenomenological
purpose, such as  muon $g-2$ anomaly ~\cite{Ma:2001md, Baek:2001kca,Heeck:2011wj, Harigaya:2013twa, Altmannshofer:2016brv, Banerjee:2018eaf,Biswas:2016yan, Biswas:2016yjr,Zhou:2022cql}, 
dark matter (DM) puzzle~\cite{Biswas:2016yan, Biswas:2016yjr,Zhou:2022cql,Costa:2022oaa,Chun:2018ibr,Baek:2008nz, Das:2013jca, Patra:2016shz,
 Biswas:2017ait, Foldenauer:2018zrz,Okada:2019sbb}, and $b \to s\mu^+\mu^-$ anomaly \cite{bs1,bs2,bs3,bs4,bs5,bs6,bs7,bs8,bs9,bs10,bs11,bs12} etc.

In this paper, we will combine the muon $g-2$ anomaly and DM observables,
 and examine the $U(1)_{L_\mu-L_\tau}$ breaking phase transition (PT), gravitational wave (GW) signatures, and the exclusion of the LHC direct searches in
an extra $U(1)_{L_\mu-L_\tau}$ gauge symmetry extension of SM. 
The model was proposed by one of our authors in \cite{Zhou:2022cql}, 
in which the new particles include vector-like leptons ($E_1,~ E_2,~ N$), $U(1)_{L_\mu-L_\tau}$ breaking scalar $S$ and gauge boson $Z'$, as well as the dark
matter candidate $X_I$ and its heavy partner $X_R$.
When the PT is first-order,
GW could be generated and detected in current and future GW
experiments, such as LISA
\cite{lisa}, Taiji \cite{taiji}, TianQin \cite{tianqin}, Big Bang Observer (BBO) \cite{bbodecigo}, DECi-hertz Interferometer
GW Observatory (DECIGO) \cite{bbodecigo} and Ultimate-DECIGO (U-DECIGO) \cite{udecigo}. 
In addition, the null results of the LHC direct searches could exclude some parameter space achieving a first-order PT (FOPT).

\section{the model}
Under the local $U(1)_{L_\mu-L_\tau}$, the second (third) generation left-handed lepton doublet and right-handed singlet, $L_\mu$, $\mu_R$ , ($L_\tau$, $\tau_R$), are
charged  with charge 1 (-1). To obtain the mass of $U(1)_{L_\mu-L_\tau}$ gauge boson $Z'$, a complex singlet scalar ${\cal S}$ is required to break the
the $U(1)_{L_\mu-L_\tau}$ symmetry. Another complex singlet scalar $X$ with $U(1)_{L_\mu-L_\tau}$ charge is introduced whose lighter component may be as 
a candidate of DM. Also we add vector-like lepton doublet fields ($E^"_{L,R}$) and singlet fields ($E'_{L,R} $) which mediate the $X$ interactions to the muon lepton,
and contribute to the muon $g-2$.
 The quantum numbers of these field under the gauge group 
$SU(3)_C\times SU(2)_L\times U(1)_Y\times U(1)_{L_\mu-L_\tau}$
are shown in Table \ref{tabquantum}.

\begin{table}[t]
  \centering
  \caption{The $U(1)_{L_\mu-L_\tau}$ quantum numbers of the new fields.}
  \label{tabquantum}
 \begin{tabular}{ccccc}
  \hline
            & SU(3)$_c$~~ & SU(2)$_L$ ~~& U(1)$_Y$ ~~& U(1)$_{L_\mu-L_\tau}$  \\ \hline
  $ X $    & {\bf 1} & {\bf 1}& $0$ & $q_x$  \\
  ${\cal S}$    & {\bf 1} & {\bf 1}& $0$ & $-2q_x$  \\
  $ E^"_{L,R} $    & {\bf 1}   & {\bf 2}& $-1/2$ & $1-q_x$  \\
  $ E'_{L,R} $    & {\bf 1} & {\bf 1}& $-1$ & $1-q_x$  \\ 
    \end{tabular}
\end{table}
The  Lagrangian is written as 
%%1901.14761
\begin{align}
  {\cal L} &= {\cal L}_{SM} -{1 \over 4} Z'_{\mu\nu} Z^{\prime\mu\nu}
              + g_{Z'} Z'^{\mu}(\bar{\mu}\gamma_\mu \mu + \bar{\nu}_{\mu_L}\gamma_\mu\nu_{\mu_L} - \bar{\tau}\gamma_\mu \tau - \bar{\nu}_{\tau_L}
             \gamma_\mu\nu_{\tau_L})\nonumber\\
            &+ \bar{E"} (i \centernot D ) E"+ \bar{E'} (i \centernot D ) E'+ (D_\mu X^\dagger) (D^\mu X)  + (D_\mu {\cal S}^\dagger) (D^\mu {\cal S}) \nonumber\\
            &-V  + \mathcal{L}_{\rm Y}.
\label{eq:model}
\end{align} 
Where  $Z'_{\mu\nu}=\partial_\mu Z'_\nu-\partial_\nu Z'_\mu$ is the field strength tensor, $D_\mu$ is the covariant derivative, and $g_{Z'}$ is the gauge coupling 
constant of the $U(1)_{L_\mu-L_\tau}$. $V$ and $\mathcal{L}_{\rm Y}$ denote the scalar potential and Yukawa interactions.

The scalar potential $V$ containing the SM Higgs parts can be given by
\begin{eqnarray} \label{scalarV} V &=& -\mu_{h}^2
(H^{\dagger} H) - \mu_{S}^2 ({\cal S}^{\dagger} {\cal S}) + m_X^2 (X^{\dagger} X) + \left[\mu X^2 {\cal S} + \rm h.c.\right]\nonumber \\
&&+ \lambda_H (H^{\dagger} H)^2 +
\lambda_S ({\cal S}^{\dagger} {\cal S})^2 + \lambda_X (X^{\dagger} X)^2 + \lambda_{SX}
({\cal S}^{\dagger} {\cal S})(X^{\dagger} X) \nonumber \\
&&+ \lambda_{HS}(H^{\dagger} H)({\cal S}^{\dagger} {\cal S}) + \lambda_{HX}(H^{\dagger} H)(X^{\dagger} X)
\end{eqnarray}
with
\begin{equation}
H=\left(\begin{array}{c} G^+ \\
\frac{1}{\sqrt{2}}\,(h_1+v_h+iG)
\end{array}\right)\,,
{\cal S}={1\over \sqrt{2}} \left( h_2+v_S+i\omega\right) \,,
X={1\over \sqrt{2}} \left( X_R+iX_I\right) \,.
\end{equation}
Where $v_h=246$ GeV and $v_S$ are respectively vacuum expectation values (VeVs) of $H$ and ${\cal S}$, 
and the $X$ field has no VeV.
One can determine the mass parameters $\mu^{2}_{h}$ and $\mu^{2}_{S}$ of Eq. (\ref{scalarV}) by the potential  minimization conditions,
\beq
\begin{split}
&\quad \mu_{h}^2 = \lam_H v_h^2 + {1 \over 2} \lam_{HS} v_S^2,\\
&\quad \mu_{S}^2 = \lam_S v_S^2 + {1 \over 2} \lam_{HS} v_h^2.\\
\end{split}
\label{min_cond}
\eeq

After ${\cal S}$ acquires the VeV, the $\mu$ term makes the complex scalar $X$ split into two real scalar fields ($X_R$, $X_I$), and
their masses are 
\begin{align}
 &m_{X_R}^2 = m_X^2 +  {1 \over 2} \lambda_{HX} v_H^2 + {1 \over 2}\lambda_{SX} v_S^2  + \sqrt{2} \mu v_S\nonumber\\
 &m_{X_I}^2 = m_X^2 +  {1 \over 2} \lambda_{HX} v_H^2 + {1 \over 2}\lambda_{SX} v_S^2  - \sqrt{2} \mu v_S.
\end{align}
After the $U(1)_{L_\mu-L_\tau}$ is broken, there is still remnant $Z_2$ symmetry, which guarantee
the lightest component $X_I$ to be as a candidate of DM.

The $\lambda_{HS}$ and $\lambda_{HX}$ terms will lead to the couplings of 125 GeV Higgs ($h$) and DM.
To suppress the stringent constraints from the DM direct detection and indirect detection 
experiments, we simply assume that the $hX_IX_I$ coupling is absent, namely taking $\lambda_{HX}=0$ and $\lambda_{HS}=0$.
Thus, the 125 GeV Higgs $h$ is purely from $h_1$ and extra CP-even Higgs $S$ is purely from $h_2$. Their masses are
given by 
\beq
m_h^2=2\lambda_H v_h^2,~~~m_S^2=2\lambda_S v_S^2.
\eeq

After ${\cal S}$ acquires VeV, the $U(1)_{L_\mu-L_\tau}$ gauge boson $Z'$ obtains a mass,
\beq m_Z' = 2g_Z' \mid q_x\mid v_S.
\eeq

The Yukawa interactions $\mathcal{L}_{\rm Y}$ can be written as
\begin{align}
-\mathcal{L}_{\rm Y,mass}
=&  m_1 \overline{E'_L} E'_R  + m_2 \overline{E^"_L} E^"_R + \kappa_{1} \overline{\mu_R} X E'_L+ \kappa_{2} \overline{L_\mu} X E^"_R \nn \\
  & + \sqrt{2}y_1 \overline{E^"_L} H E'_R + \sqrt{2}y_2 \overline{E^"_R} H E'_L +\frac{\sqrt{2}m_\mu}{v} \overline{L_\mu}H \mu_R
+ {\rm h.c.}.
\label{lag-yuka}
\end{align} 
After the Electroweak symmetry breaking, the vector-like lepton masses are given by
\begin{align}
 M_E = 
\begin{pmatrix}
 m_1 & y_2 v_h \\ y_1 v_h & m_2
\end{pmatrix}.   
\end{align}
By making a bi-unitary transformation with the rotation matrices for the right-handed
fields and the left-handed fields, 
\begin{align}
 U_L = 
\begin{pmatrix}
 c_L & -s_L \\ s_L & c_L
\end{pmatrix},
\quad 
 U_R = 
\begin{pmatrix}
 c_R & -s_R \\ s_R & c_R
\end{pmatrix},
\end{align}
where $c_{L,R}^2 + s_{L,R}^2 = 1$,
we can diagonalize the mass matrix for the vector-like lepton,
\begin{align}
U_L^\dag M_E U_R = \mathrm{diag}\left(m_{E_1}, m_{E_2}\right).
\end{align}
The $E_1$ and $E_2$ are the 
mass eigenstates of charged vector-like leptons, and the mass of neutral vector-like lepton $N$ is
\beq\label{mnn}
m_{N}=m_2=m_{E_2} c_L c_R+m_{E_1} s_L s_R.
\eeq
The $E_1$ and $E_2$ can mediate $X_R$ and $X_I$ interactions to muon lepton,
\beq
-\mathcal{L}_{\rm X} \supset \frac{1}{\sqrt{2}}(X_R+ i X_I)\left[\bar{\mu}_R (\kappa_1 c_L E_{1L} -\kappa_1 s_L E_{2L}) + \bar{\mu}_L (\kappa_2 s_R E_{1R} + \kappa_2 c_R E_{2R})\right] + h.c.~,
\eeq
and have the couplings to the 125 GeV Higgs,
\bea
-\mathcal{L}_{\rm h} &\supset& \frac{m_{E_1}(c_L^2 s_R^2 +c_R^2 s_L^2)-2m_{E_2}s_L c_L s_R c_R}{v_h}~ h\bar{E}_1E_1,\nonumber\\
&&+\frac{m_{E_2}(s_L^2 c_R^2 +c_L^2 s_R^2)-2m_{E_1}s_L c_L s_R c_R}{v_h}~ h\bar{E}_2E_2.\label{yuka-he1e1}
\eea

\section{Dark matter and muon $g-2$}
We fix on $m_h=$ 125 GeV, $v_h=$ 246 GeV, $q_x=-1$, $\lambda_{HS}=0$, and
$\lambda_{HX}=0$, and take
$g_{Z'}$, $m_{Z'}$, $\lambda_X$, $\lambda_{SX}$, $m_S$, $m_{X_R}$, $m_{X_I}$, $m_{E_1}$,
$m_{E_2}$, $s_L$, $s_R$, $\kappa_1$, and $\kappa_2$ as the free parameters.
To maintain the perturbativity, we conservatively choose
\bea
&&\mid\lambda_{SX}\mid\leq 4\pi,~~\mid\lambda_X\mid\leq 4\pi,\nonumber\\
&&-\frac{1}{2}\leq\kappa_1 \leq \frac{1}{2},~~-\frac{1}{2}\leq\kappa_2 \leq \frac{1}{2},
\eea
and take the mixing parameters $s_L$ and $s_R$ as
\beq
-\frac{1}{\sqrt{2}} \leq s_L \leq \frac{1}{\sqrt{2}}, ~~-\frac{1}{\sqrt{2}} \leq s_R \leq \frac{1}{\sqrt{2}}.
\eeq
The mass parameters are scanned over in the following ranges:
\begin{align}
 & 60~ {\rm GeV} \leq m_{X_I} \leq 500~ {\rm GeV},~~~ m_{X_I} \leq m_{X_R} \leq 500 ~{\rm GeV},\nonumber\\
 &m_{X_I} \leq m_{E_1} \leq 500 ~{\rm GeV},~~~m_{X_I} \leq m_{E_2} \leq 500 ~{\rm GeV},\nonumber\\
 & 100 ~{\rm GeV} \leq m_{Z'} \leq 500 ~{\rm GeV},~~~100 ~{\rm GeV} \leq m_{S} \leq 500 ~{\rm GeV}.
\end{align}
The mass of neutral vector-like lepton $N$ is a function of $m_{E_1}$, $m_{E_2}$, $s_L$ and $s_R$, and $m_{X_I}<m_N$ is imposed.
We require 0 $<g_{Z'}/m_{Z'}\leq$ (550 GeV)$^{-1}$ to be consistent with the bound of the neutrino trident process \cite{trident}.

The potential stability in Eq. (\ref{scalarV}) requires the following condition,
\begin{eqnarray}
&&\lambda_H \geq 0 \,, \quad \lambda_S \geq 0 \,,\quad \lambda_X \geq 0 \,,\quad \nonumber \\
&& \lambda_{HS} \geq - 2\sqrt{\lambda_H \,\lambda_{S}}  \,, \quad
\lambda_{HX} \geq - 2\sqrt{\lambda_H \,\lambda_{X}}  \,, \quad
\lambda_{SX} \geq - 2\sqrt{\lambda_S \,\lambda_{X}} \,, \quad\nonumber \\
&&\sqrt{\lambda_{HS}+2\sqrt{\lambda_H \,\lambda_S}}~\sqrt{\lambda_{HX}+
2\sqrt{\lambda_H \,\lambda_{X}}}
~\sqrt{\lambda_{SX}+2\sqrt{\lambda_S\,\lambda_{X}}} \nonumber \\
&&+ 2\,\sqrt{\lambda_H \lambda_S \lambda_{X}} + \lambda_{HS} \sqrt{\lambda_{X}}
+ \lambda_{HX} \sqrt{\lambda_S} + \lambda_{SX} \sqrt{\lambda_H} \geq 0 \,.
\end{eqnarray}

The one-loop diagrams with the vector-like lepton can give additional corrections
to the oblique parameters ($S,~T,~U$), which can be calculated as in Refs. \cite{stu,mc-stu,1305.4712}.
Taking the recent fit results of Ref. \cite{pdg2020}, we use the following 
values of $S$, $T$, and $U$,
\beq
S=-0.01\pm 0.10,~~  T=0.03\pm 0.12,~~ U=0.02 \pm 0.11, 
\eeq
with the correlation coefficients 
\beq
\rho_{ST} = 0.92,~~  \rho_{SU} = -0.80,~~  \rho_{TU} = -0.93.
\eeq
Also the one-loop diagrams of the charged vector-like leptons $E_1$ and $E_2$ can contribute to the $h\to \gamma\gamma$ decay, and
the bound of the diphoton signal strength of the 125 GeV Higgs is imposed \cite{pdg2020}, 
\beq
\mu_{\gamma\gamma}= 1.11^{+0.10}_{-0.09}.
\eeq

In the model, the dominant corrections to the muon $g-2$ are from the one-loop diagrams with the vector-like leptons ($E_1,~ E_2)$ and scalar fields ($X_R$ and $X_I$),
which are approximately calculated as in Refs. \cite{1305.3522,1906.11297,0902.3360}
\bea
\Delta a_\mu &=&\frac{1}{32\pi^2}m_\mu \left(\kappa_1 c_L \kappa_2 s_R H(m_{E_1},m_{X_R})-\kappa_1 s_L \kappa_2 c_R H(m_{E_2},m_{X_R}) \nonumber \right.\\
&&\left. +\kappa_1 c_L \kappa_2 s_R H(m_{E_1},m_{X_I})-\kappa_1 s_L \kappa_2 c_R H(m_{E_2},m_{X_I}) \right),\label{eq-mug2}
\eea
where the function 
\beq
H(m_{f},m_{\phi})=\frac{m_f}{m_{\phi}^2}\frac{(r^2-4r+2\log{r}+3)}{(r-1)^3}
\eeq
with $r=\frac{m_f^2}{m_{\phi}^2}$.
The combined average for the muon $g-2$ with Fermilab E989 \cite{fermig2} and Brookhaven E821 \cite{E821},
the difference from the SM prediction becomes
\bea
\Delta a_\mu=a_\mu^{exp}-a_\mu^{SM}=(25.1\pm5.9)\times10^{-10},
\eea 
which shows $4.2\sigma$ discrepancy from the SM. 
Very recently, on August 10, 2023, the E989 experiment at Fermilab released an update regarding the measurement from Run-2 and Run-3 \cite{mug2-2023}. 
The new combined value yields a deviation of
\bea
\Delta a_\mu=a_\mu^{exp}-a_\mu^{SM}=(24.9\pm4.8)\times10^{-10},
\eea 
which leads to a $5.1\sigma$ discrepancy.
Whereas the recent lattice calculation \cite{mug2-lat} and the experiment determination \cite{mug2-had} of the hadron
vacuum polarization contribution to the muon $g-2$ point the value closer to the SM prediction,
and hence the tension relaxes to a few sigma level.

If kinematically allowed, the DM pair-annihilation processes includes $X_IX_I \to \mu^+\mu^-,~Z'Z',~SS$. In addition, a small 
mass splitting between the DM and the other new particles ($E_1,~E_2,~N$,~$X_R$) can lead to coannihilation.
The Planck collaboration reported the relic density of cold DM in the universe,
 $\Omega_{c}h^2 = 0.1198 \pm 0.0015$ \cite{planck},
and the theoretical prediction of the model is calculated by $\textsf{micrOMEGAs-5.2.13}$ \cite{micomega}.

%%%%%%%%%%%%%%%%%%%%%
\begin{figure}[tb]
%\begin{center}
 \epsfig{file=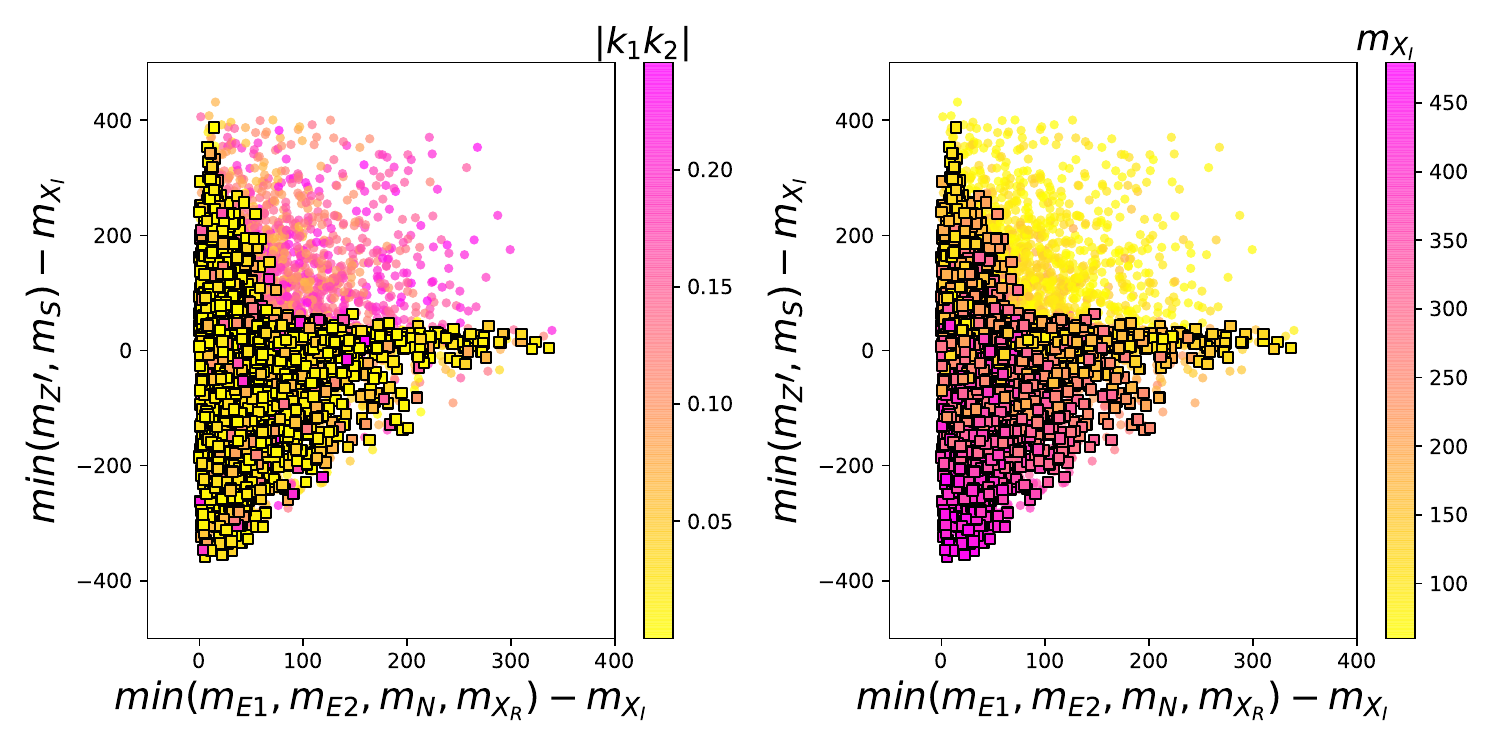,height=8.0cm}
 %\end{center}
\vspace{-0.3cm} \caption{The surviving samples explaining the DM relic density and muon $g-2$ anomaly while satisfying the constraints from theory, 
oblique parameter, and 125 GeV Higgs diphoton signal. The circles and squares are excluded and allowed by the DM relic data.} \label{figdmg2}
\end{figure}
%%%%%%%%%%%%%%%%%%%%

After imposing the constraints of theory, oblique parameters, and the diphoton signal data of the 125 GeV Higgs, we project
the samples accommodating the DM relic density and muon $g-2$ anomaly within $2\sigma$ ranges in Fig. \ref{figdmg2}. 
From Fig. \ref{figdmg2} we find that the correct DM relic density can be obtained for most of the parameter region 
of $min(m_{E_1},~m_{E_2},~m_N,~m_{X_R})-m_{X_I} < 350$ GeV and -400 GeV $<min(m_{Z'},m_S) - m_{X_I}<$ 400 GeV. 
The DM relic density is mainly produced via the DM pair-annihilation processes $X_IX_I \to ~Z'Z',~SS$ for the region of $min(m_{Z'},m_S) < m_{X_I}$, 
the coannihilation processes for the region of $min(m_{E_1},~m_{E_2},~m_N,~m_{X_R})$ close to $m_{X_I}$, and 
$X_IX_I \to \mu^+\mu^-$ for the region of both $min(m_{E_1},~m_{E_2},~m_N,~m_{X_R})>m_{X_I}$ and $min(m_{Z'},m_S) > m_{X_I}$.
 However, once the explanation of muon $g-2$ anomaly is simultaneously required, most of region of $min(m_{E_1},~m_{E_2},~m_N,~m_{X_R})>m_{X_I}$ and $min(m_{Z'},m_S) > m_{X_I}$
is ruled out. This is because the muon $g-2$ anomaly favors small interactions between the vector-like leptons and muon mediated by $X_I$, which leads to
$X_IX_I \to \mu^+\mu^-$ process to fail to produce the correct DM relic density.

The $X_I$ does not couple to the SM quark, and its couplings to the muon lepton and vector-like leptons are restricted by the muon $g-2$ anomaly.
Therefore, the model can accommodate the bound from the DM direct detection naturally.

\section{$U(1)_{L_\mu-L_\tau}$ breaking phase transition}

At high temperatures, the global minimum of the finite-temperature effective potential
is at the origin, i.e. $SU(2)_L\times U(1)_Y\times U(1)_{L_\mu-L_\tau}$ is unbroken. When the temperature drops,
the potential changes and at some point develops a minimum at non-vanishing field values.
The PT between the unbroken and the broken phase can proceed in
basically two different ways. In a FOPT, at the critical temperature $T_C$, the two degenerate minima 
will be at different points in field space, typically with a potential barrier in between. For
a second-order (cross-over) transition, the broken and symmetric minimum are not degenerate 
until they are at the same point in field space. In this paper we focus on a first-order $U(1)_{L_\mu-L_\tau}$ breaking PT.

\subsection{The thermal effective potential}
In order to examine PT, we first take $h_1$, $h_2$, and $X_r$ as the field configurations, and 
obtain the field dependent masses of the scalars ($h$, $S$, $X_R$, $X_I$), the Goldstone boson ($G,~\omega,~G^{\pm}$), the gauge boson, and fermions.
The field dependent masses of scalars are given 
\begin{align}
\hm^2_{h,S,X_R} &=\rm{eigenvalues} ( \widehat{\mathcal{M}^2_P} ) \ , \\
\hm^2_{G,\omega,X_I} &=\rm{eigenvalues} ( \widehat{\mathcal{M}^2_A}) \ , \\
\hm^2_{G^\pm} &=\lam_H h_1^2 -\lam_H v_h^2  \ ,
\end{align}
with
\begin{align}
\widehat{\mathcal{M}^2_P}_{11} &=3\lam_{H} h^2_1 -\lam_{H} v_h^2\nonumber\\
\widehat{\mathcal{M}^2_P}_{22} &=-\lam_S v_S^2 + 3 \lam_S h_2^2 + {\lam_{SX} \over 2} X_r^2\nonumber\\
\widehat{\mathcal{M}^2_P}_{33} &=m_X^2+\sqrt{2}\mu h_2 + {\lam_{SX} \over 2} h_2^2 + 3\lam_X X_r^2\nonumber\\
\widehat{\mathcal{M}^2_P}_{23} &=\widehat{\mathcal{M}^2_P}_{32}=\sqrt{2} \mu X_r + \lam_{SX} h_2 X_r \nonumber\\
\widehat{\mathcal{M}^2_P}_{12} &=\widehat{\mathcal{M}^2_P}_{21}=\widehat{\mathcal{M}^2_P}_{13} =\widehat{\mathcal{M}^2_P}_{31}=0\nonumber\\
\widehat{\mathcal{M}^2_A}_{11} &=\lam_{H} h^2_1 -\lam_{H} v_h^2\nonumber\\
\widehat{\mathcal{M}^2_A}_{22} &=-\lam_{S} v_S^2 + \lam_S h^{2}_{2}+ {\lam_{SX} \over 2} X_r^{2}\nonumber\\
\widehat{\mathcal{M}^2_A}_{33} &=m_X^2-\sqrt{2} \mu h_2 + \lam_X X^{2}_{r}+ {\lam_{SX} \over 2} h_2^{2}\nonumber\\
\widehat{\mathcal{M}^2_A}_{23} &=\widehat{\mathcal{M}^2_A}_{32}=-\sqrt{2} \mu X_r\nonumber\\
\widehat{\mathcal{M}^2_A}_{12} &=\widehat{\mathcal{M}^2_A}_{21}=\widehat{\mathcal{M}^2_A}_{13}=\widehat{\mathcal{M}^2_A}_{31}=0.
\end{align}

The field dependent masses of gauge boson are
\begin{align}
%\hm^2_t &= {1\over 2} y^2_t \left(h^{2}_{1} +h^{2}_{2} \right) \ , \\
&\hm^2_{W^\pm} = {1\over 4} g^2 h^{2}_{1},
&  \hm^2_{Z} = {1\over 4} (g^2+g'^2) h^{2}_{1}, \\
&\quad \hm^2_{\gamma}=0 ,
& \hm^2_{Z'} = q_x^2 g_{Z'}^2 (4h^{2}_{2}+X^2_{r}), 
\end{align}

The field dependent masses of vector-like lepton are
\begin{align}
\widehat{\mathcal{M}}^2_{E_1,E_2,\mu} &=\rm{eigenvalues} ( \widehat{\mathcal{M}_E}\widehat{\mathcal{M}_E}^T )
\end{align}
with 
\bea
&& \widehat{\mathcal{M}_E}= \begin{pmatrix} m_1 & \quad y_2 h_{1} & \quad \kappa_1 X_r\\[5pt]
y_1 h_{1} & \quad   m_2 & \quad 0\\[5pt] 0 & \quad \kappa_2 X_r & \quad \frac{m_{\mu} h_1}{v_h} \end{pmatrix} .\label{m-cp-odd}
\eea

For the quarks of SM, we only consider the top quark,
\begin{align}
\widehat{\mathcal{M}^2_{t}} &=y_t^2 h_1^2
\end{align}
with $y_t=m_t/v_h$.

In order to examine the $U(1)_{L_\mu-L_\tau}$ breaking PT, we need to study the thermal effective potential $V_{eff}$ in terms of the classical fields ($h_1,h_2,X_r$),
which is composed of four parts:
\begin{align}
V_{eff} (h_1,h_2,X_r,T)= &V_{0}(h_1,h_2,X_r) + V_{CW}(h_1,h_2,X_r) + V_{CT}(h_1,h_2,X_r) \nonumber\\
&+ V_{T}(h_1,h_2,X_r,T) + V_{ring}(h_1,h_2,X_r,T).
\label{veff0}
\end{align}
 $V_{0}$ is the tree-level potential, $V_{CW}$ is the Coleman-Weinberg (CW) potential \cite{vcw}, 
$V_{CT}$ is the counter term, $V_{T}$ is the thermal correction \cite{vloop}, and $V_{ring}$ is the resummed daisy corrections \cite{vring1,vring2}.
In this paper, we calculate $V_{eff}$ in the Landau gauge.
%1809.09110

The tree-level potential $V_0$ in terms of their classical fields ($h_1,h_2,X_r$) 
from the Eq. (\ref{scalarV}), 
\bea
V_0(h_1,h_2,X_r)&=& -\frac{\mu_h^2}{2} h_1^2 - \frac{\mu_S^2}{2} h_2^2 + \frac{m_X^2}{2} X_r^2 + \frac{\mu}{\sqrt{2}} h_2 X_r^2\nonumber\\
&&+\frac{\lambda_H}{4} h_1^4 + \frac{\lambda_S}{4} h_2^4 + \frac{\lambda_X}{4} X_r^4 + \frac{\lambda_{SX}}{4} X_r^2 h_2^2 .
\eea

The CW potential in the $\overline{\rm MS}$ scheme at 1-loop level has the form \cite{vcw}:
\beq
\label{eq:CWpot}
V_{\rm CW}(h_{1},h_{2},X_r) = \sum_{i} (-1)^{2s_i} n_i\frac{\hm_i^4 (h_{1},h_{2},X_r)}{64\pi^2}\left[\ln \frac{\hm_i^2 (h_{1},h_{2},X_r)}{Q^2}-C_i\right] \; \ ,
\end{equation}
where $i=h,S,X_R,X_I,G,\omega,G^\pm,W^\pm,Z,Z',t,E_1,E_2,\mu$, and $s_i$ is the spin of particle i. $Q$ is a renormalization scale, and we take $Q=m_S$.
The constants $C_i =\frac{3}{2}$ for scalars or fermions and
$C_i = \frac{5}{6}$ for gauge bosons.
$n_i$ is the number of degree of freedom,
\begin{align}
&n_h=n_S=n_{X_R}=n_{X_I}=n_G=n_{\omega}=1\nonumber\\ 
&n_{G^\pm}=2,~~n_{W^\pm}=6,~n_{Z}=n_{Z'}=3\nonumber\\
&n_{t}=12,~~n_{E_1}=n_{E_2}=n_{\mu}=4.
\end{align}

With $V_{CW}$ being included in the potential, the minimization conditions of scalar potential 
 and the CP-even mass matrix will be shifted
slightly. To maintain these relations, the counter terms $V_{ct}$ should be
added,
\begin{align}
\label{eq:Vct}
V_{\rm CT}&=\delta m_1^2 h_{1}^2+\delta m_2^2 h_{2}^2 + \delta m_X^2 X_r^2 + \delta \lam_H h_{1}^4+\delta \lam_{S} h_{2}^4 +\delta \lam_X X_{r}^4\nonumber\\
& + \delta \mu X_r^2 h_2 +\delta \lam_{SX} h_2^2 X_r^2.
\end{align}
The relevant coefficients are determined by
\beq
\label{eq:V1der}
\frac{\partial V_{\rm CT}}{\partial h_{1}} = -\frac{\partial V_{\rm CW}}{\partial h_{1}}\;, \quad \frac{\partial V_{\rm CT}}{\partial h_{2}} = -\frac{\partial V_{\rm CW}}{\partial h_{2}},\; \quad \frac{\partial V_{\rm CT}}{\partial X_r} = -\frac{\partial V_{\rm CW}}{\partial X_r},
\eeq
\begin{align}
\label{eq:V2der}
\frac{\partial^{2} V_{\rm CT}}{\partial h_{1}\partial h_{1}} &= - \frac{\partial^{2} V_{\rm CW}}{\partial h_{1}\partial h_{1}}\;, \quad
\frac{\partial^{2} V_{\rm CT}}{\partial h_{2}\partial h_{2}} = - \frac{\partial^{2} V_{\rm CW}}{\partial h_{2}\partial h_{2}}\;, \quad
\frac{\partial^{2} V_{\rm CT}}{\partial X_{r}\partial X_{r}} = - \frac{\partial^{2} V_{\rm CW}}{\partial X_{r}\partial X_{r}}\;,\nonumber\\
\frac{\partial^{2} V_{\rm CT}}{\partial h_{1}\partial h_{2}} &= - \frac{\partial^{2} V_{\rm CW}}{\partial h_{1}\partial h_{2}}\;,  \quad
\frac{\partial^{2} V_{\rm CT}}{\partial h_1\partial X_r} = - \frac{\partial^{2} V_{\rm CW}}{\partial h_1\partial X_r}\;, \quad
\frac{\partial^{2} V_{\rm CT}}{\partial h_2\partial X_r} = - \frac{\partial^{2} V_{\rm CW}}{\partial h_2\partial X_r}\;, \quad
\end{align}
%\begin{align}
%\label{eq:V2der}
%\frac{\partial^{2} V_{\rm CT}}{\partial h_{1}\partial h_{1}} &= - \frac{\partial^{2} V_{\rm CW}}{\partial h_{1}\partial h_{1}}\;, \quad
%\frac{\partial^{2} V_{\rm CT}}{\partial h_{2}\partial h_{2}} = - \frac{\partial^{2} V_{\rm CW}}{\partial h_{2}\partial h_{2}}\;, \nonumber\\
%\frac{\partial^{2} V_{\rm CT}}{\partial X_{r}\partial X_{r}} &= - \frac{\partial^{2} V_{\rm CW}}{\partial X_{r}\partial X_{r}}\;, 
%\frac{\partial^{2} V_{\rm CT}}{\partial h_2\partial X_r} = - \frac{\partial^{2} V_{\rm CW}}{\partial h_2\partial X_r}\;, \quad
%\end{align}
which are evaluated at the EW minimum of $\{ h_{1}=v_h, h_{2}=v_S, X_r=0 \}$ on both sides.
As a result, the VeVs of $h_{1}$, $h_{2}$, $X_r$ and the CP-even mass matrix will not be shifted.
We check that the following relations hold true
\begin{align}
\label{eq:V2der}
\frac{\partial^{2} V_{\rm CW}}{\partial h_{1}\partial h_{2}} =0\;, \quad
\frac{\partial^{2} V_{\rm CW}}{\partial h_{1}\partial X_{r}} =0\;.
\end{align}
For the left seven equations, there are eight parameters to be fixed, so that one
renormalization constant is left for determination. We choose to use 
$\delta m_1^2$, $\delta m_2^2$, $\delta m_X^2$, $\delta \lam_H$, $\delta \lam_{S}$, $\delta \lam_X$, $\delta \lam_{SX}$,
and set $\delta \mu=0$. 
It is a well-known problem that the second derivative of the
CW potential in the vacuum suffers from logarithmic divergences originating
from the vanishing Goldstone masses. In order to avoid the
problems with infrared divergent Goldstone contributions, we simply remove the
Goldstone corrections in the renormalization conditions following the approaches of \cite
{1107.3559,Espinosa:2011ax}. This is simply a change of renormalization conditions and the shift it
causes in the potential shape is negligible.

The thermal contributions $V_T$ to the potential can be written as \cite{vloop}
\beq
\label{potVth}
 V_{\rm th}(h_{1},h_{2},X_r,T) = \frac{T^4}{2\pi^2}\, \sum_i n_i J_{B,F}\left( \frac{ \hm_i^2(h_{1},h_{2},X_r)}{T^2}\right)\;,
\eeq
where $i=h,S,X_R,X_I,G,\omega,G^\pm,W^\pm,Z,Z',t,E_1,E_2,\mu$, and the functions $J_{B,F}$ are 
\beq
\label{eq:jfunc}
J_{B,F}(y) = \pm \int_0^\infty\, dx\, x^2\, \ln\left[1\mp {\rm exp}\left(-\sqrt{x^2+y}\right)\right].
\eeq

Finally, the thermal corrections with resumed ring diagrams are given \cite{vring1,vring2}.
\beq
V_{\rm ring}\left(h_{1},h_{2},X_r, T\right) =-\frac{T}{12\pi }\sum_{i} n_{i}\left[ \left( \bar{M}_{i}^{2}\left(h_{1},h_{2},X_r,T\right) \right)^{\frac{3}{2}}-\left( \hm_{i}^{2}\left(h_{1},h_{2},X_r,T\right) \right)^{\frac{3}{2}}\right] ,
\label{eq:daisy}
\eeq
where $i=h,S,X_R,X_I,G,\omega,G^\pm,W^\pm_L,Z_L,Z'_L,\gamma_L$. The $W^\pm_L,~Z_L,~Z'_L$, and $\gamma_L$ are the longitudinal gauge bosons with
$n_{W^\pm_L}=2,~n_{Z_L}=n_{Z'_L}=n_{\gamma_L}=1$.
The thermal Debye masses $\bar{M}_{i}^{2}\left(h_{1},h_2,X_r,T\right)$ for the CP-even and CP-odd scalar fields are the eigenvalues of the full mass matrix, 
\begin{equation}
\label{eq:thermalmass}
\bar{M}_{i}^{2}\left( h_{1},h_{2},X_r,T\right) ={\rm eigenvalues} \left[\widehat{\mathcal{M}_{X}^2}\left( h_{1},h_{2},X_r\right) +\Pi _{X}(T)\right]  ,
\end{equation}%
where $\Pi_X$ ($X=P,A$) are given by 
\begin{align}
(\Pi_{P,A})_{11} &= \left[{9g^2\over 2} + {3g'^2\over 2} + 12y_t^2  + 4(y_2^2+y_1^2) +12\lam_{H} \right] {T^2 \over 24} \nonumber\\
(\Pi_{P,A})_{22} &= \left[8\lam_S + 2\lam_{SX} + 24q_x^2 g^2_{Z'} \right] {T^2 \over 24} \nonumber\\
(\Pi_{P,A})_{33} &= \left[8\lam_X + 2\lam_{SX} + 6q_x^2 g^2_{Z'} + 2\kappa_1^2+4\kappa_2^2 \right] {T^2 \over 24} \nonumber\\
(\Pi_{P,A})_{12} &=(\Pi_{P,A})_{21} =(\Pi_{P,A})_{13} =(\Pi_{P,A})_{31}  =(\Pi_{P,A})_{32}  =(\Pi_{P,A})_{23} =0.
\end{align}
The thermal Debye mass of $G^{\pm}$ is 
\begin{align}
\label{eq:thermalmass}
\bar{M}_{G^+G^-}^{2}\left( h_{1},h_{2},X_r,T\right) =\lam_H h_1^2 -\lam_H v_h^2  +\Pi _{P11}.
\end{align}

The Debye masses of longitudinal gauge bosons are
\begin{align}
\bar{M}_{W^{\pm}_L}^2 \left( h_{1},h_{2},X_r,T\right)&= {1 \over 4} g^2 h^2_1 + {11\over 6} g^2 T^2, \nonumber\\
\bar{M}_{Z_L,\gamma_L}^{2}\left( h_{1},h_{2},X_r,T\right) &=\frac{1}{8}\left(g^2+g'^2\right)\left(h_1^2+\frac{22}{3}T^2\right)
\pm \frac{1}{2}\Delta, \nonumber\\
%\bar{M}_{Z'_L}^2\left( h_{1},h_{2},X_r,T\right)  &= q_x^2 g_{Z'}^2 (4h^{2}_{2}+X^2_{r}) + {1\over 6} g^2_{Z'} T^2 \left[6+6 (1-q_x)^2+ 10q_x^2\right], 
\bar{M}_{Z'_L}^2\left( h_{1},h_{2},X_r,T\right)  &= q_x^2 g_{Z'}^2 (4h^{2}_{2}+X^2_{r}) + {1\over 6} g^2_{Z'} T^2 \left[12-12q_x+ 16q_x^2\right], 
\end{align}
with $\Delta=\sqrt{\left[\frac{1}{4}\left(g^2-g'^2\right)\left(h_1^2+\frac{22}{3}T^2\right)\right]^2+\frac{1}{4}g^2 g'^2 h_1^2}$.

\begin{figure}[tb]
%\begin{center}
 \epsfig{file=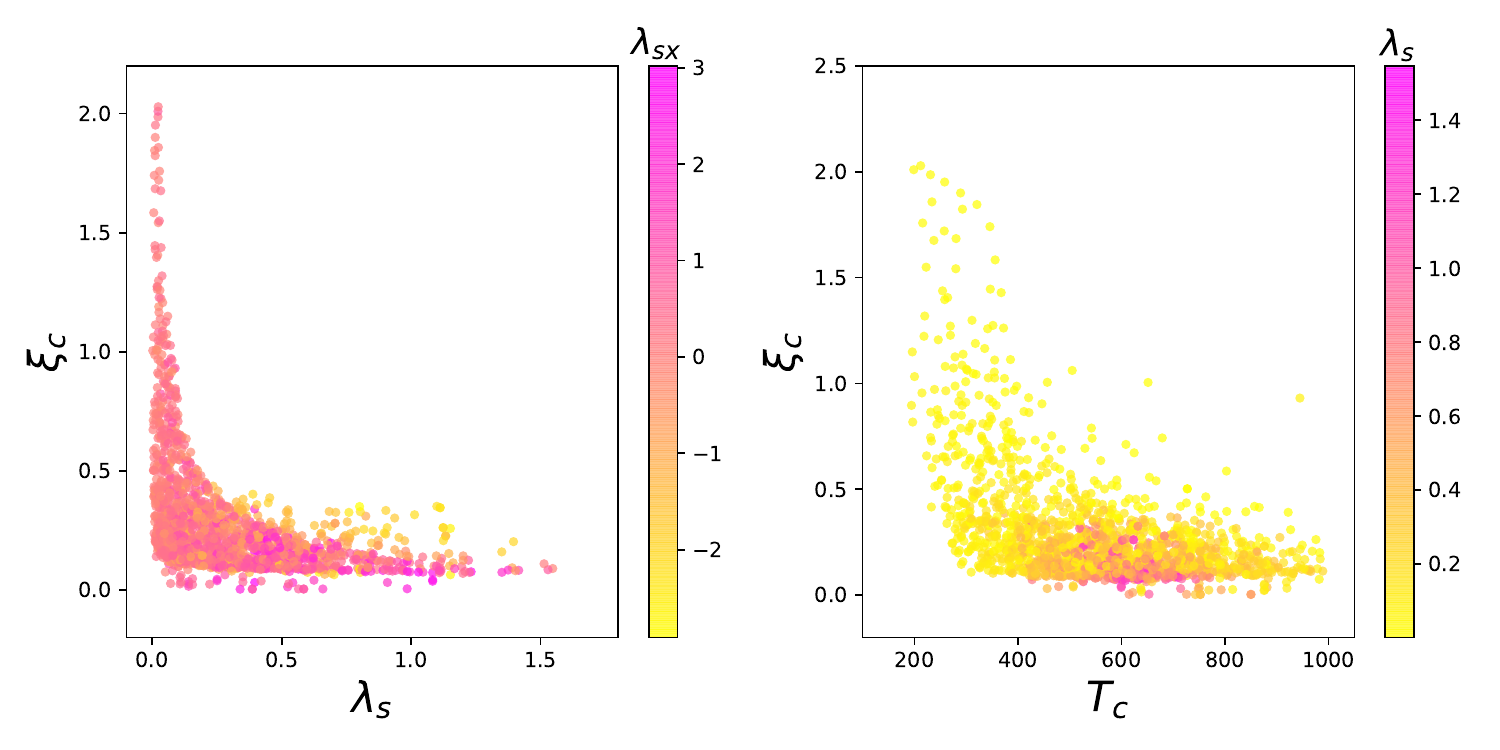,height=8.0cm}
 %\end{center}
\vspace{-0.5cm} \caption{The surviving samples achieving FOFP and accommodating the DM relic
density, muon $g-2$ anomaly, and various constraints mentioned previously. $\xi_C=\frac{<h_2>}{T_C}$ denotes
the strength of the $U(1)_{L_\mu-L_\tau}$ breaking FOPT at the critical temperature $T_C$.} \label{figpt}
\end{figure}

\begin{figure}
 \epsfig{file=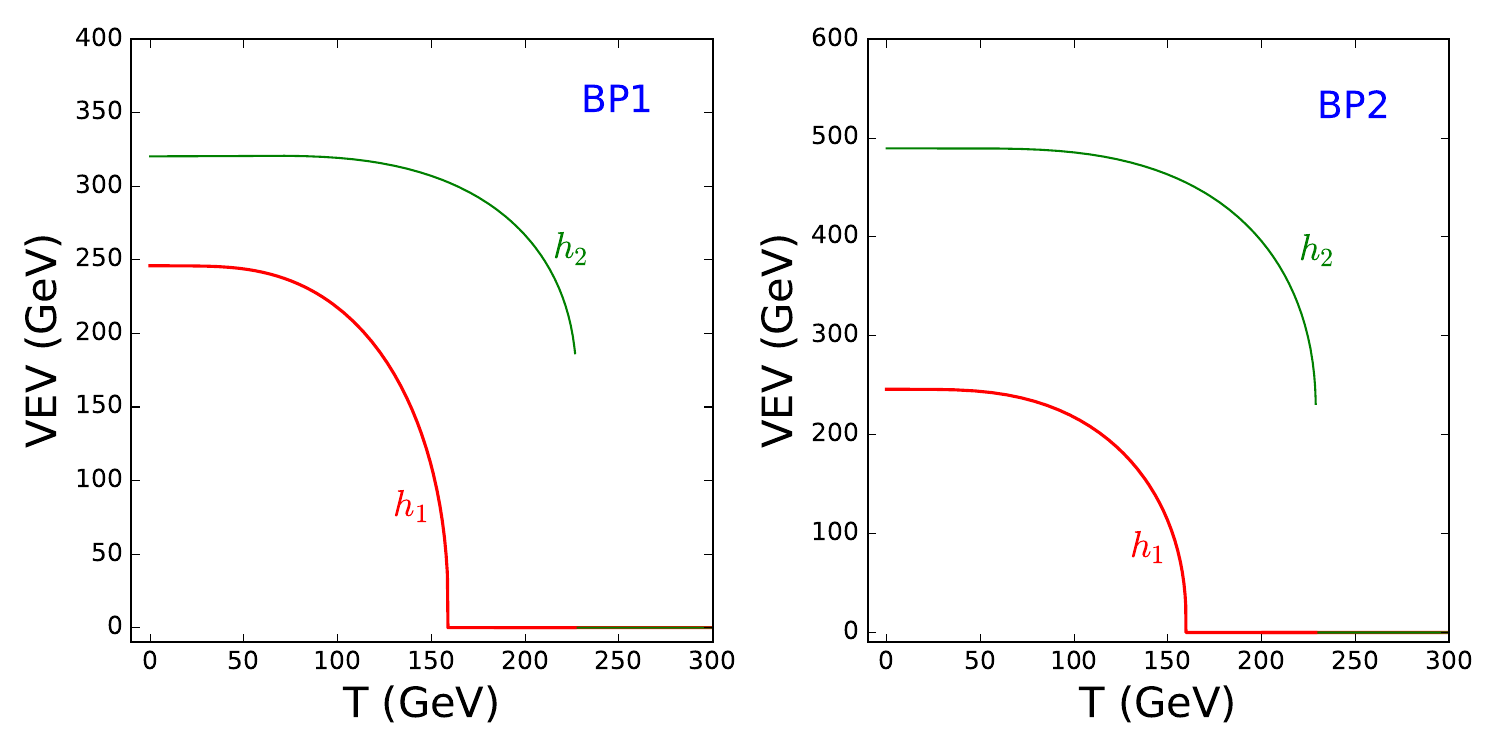,height=8.cm}
 %\end{center}
\vspace{-0.5cm} \caption{The phase histories of the BP1 and BP2. The field configuration $X_r$ is not shown as the 
minima at any temperatures locate at $\langle X_r\rangle$= 0.} 
 \label{figpt12}
\end{figure}

In Fig. \ref{figpt}, we  display the scatter plots achieving FOFP and accommodating the DM relic
density, muon $g-2$ anomaly, and various constraints mentioned previously. We observe that the strength of FOPT is sensitive to
the parameter $\lambda_S$. As $\lambda_S$ decreases, the the critical temperature $T_C$ tends to decrease, and the strength of FOPT 
tends to increase. We pick out two benchmark points (BPs) to show how the PT happens.
The input parameters of BP1 and BP2 are displayed in Table \ref{tabgrav}, and
their phase histories are exhibited in Fig.~\ref{figpt12} on field configurations versus temperature plane.
For the BP1 and BP2, the potential minima at any temperatures always locate at $\langle X_r\rangle$=0. 
As the universe cools, a FOPT takes place during which 
the $h_2$ acquires a nonzero VeV and the other two fields remain zero. As the temperature continues to decrease,
the $h_1$ starts to develop a nonzero VeV during the second PT which is second-order.
Finally, the observed vacuum is obtained at the present temperature.

%%%%%%%%%%%%%%%%%%%%
\begin{table}
\begin{footnotesize}
\begin{tabular}{| c | c | c | c |c | c | c | c |c | c | c | c |c | c | c | c |}
\hline
 &$\lambda_{SX}$ & $\lambda_{X}$ &$\lambda_{s}$ & $k_{1}k_{2}$ &$s_L$ & $s_R$ & $g_Z\prime$ & $m_{X_I}$(GeV)  \\
\hline
%%%%%%%hh
$\textbf{BP1}$   & ~~0.298~~& ~~1.478~~& ~~0.067~~ &~~ 0.0061~~& ~~0.113~~ &~~ 0.254~~& ~~0.481~~& ~~302.9~~ \\
 \hline
$\textbf{BP2}$   & ~~0.304~~& ~~0.337~~ & ~~0.017~~ &~~ -0.0164~~& ~~-0.280~~ &~~ -0.374~~& ~~0.319~~& ~~319.2~~ \\
 \hline
\end{tabular}

\begin{tabular}{| c | c | c | c |c | c | c | c |c | c | c | c |c | c | }
\hline
& $m_{X_R}$(GeV)& ~~$m_{E_1}$(GeV)~~& ~~$m_{E_2}$(GeV)~~ &~~$m_Z\prime$(GeV)~~ & ~~$m_S$(GeV)~~ \\
\hline
%%%%%%%hh
$\textbf{BP1}$   &~~ 410.2~~& ~~357.4~~ &~~454.6~~& ~~308.4~~& ~~117.1~~ \\
\hline
$\textbf{BP2}$   &~~331.4~~& ~~363.8~~ &~~469.0~~& ~~312.3~~& ~~91.6~~  \\
 \hline
\end{tabular}
%%%%%%%%%%%%%%%%%%%%%
\end{footnotesize}
\caption{Input parameters for the BP1 and BP2.}
\label{tabgrav}
\end{table}

%%%%%%%%%%%%%%%%%%%%
%%%%%%%%%%%%%%%%%%%%%
%%%%%%%%%%%%%%%%%%%%

\section{Collider and gravitational wave signature}
\subsection{Limits from the collider experiments}
Since the $U(1)_{L_\mu-L_\tau}$ gauge boson $Z'$ does not couple to
quarks, the $Z'$ is rather hard to produce at the LHC. The vector-like leptons are mainly pair produced at the LHC via the
electroweak processes mediated by the SM gauge bosons, and the $E_{1,2} \to \mu X_I$ and $N\to \nu_\mu X_I$ are main
decay modes of vector-like leptons. 
Thus, we employ the ATLAS analysis of $2\ell+E_T^{miss}$ with 139 fb$^{-1}$ integrated luminosity data to restrict our model \cite{1908.08215}, 
which is implemented in the $\textsf{MadAnalysis5}$ \cite{ma5-1,ma5-2,ma5-3} with
 assuming 95\% confidence level for the exclusion limit.
The simulations for the samples are performed by \texttt{MG5\_aMC-3.3.2}~\cite{Alwall:2014hca} 
with \texttt{PYTHIA8}~\cite{Torrielli:2010aw} and 
\texttt{Delphes-3.2.0}~\cite{deFavereau:2013fsa}.

%%%%%%%%%%%%%%%%%%%%%
\begin{figure}
 \epsfig{file=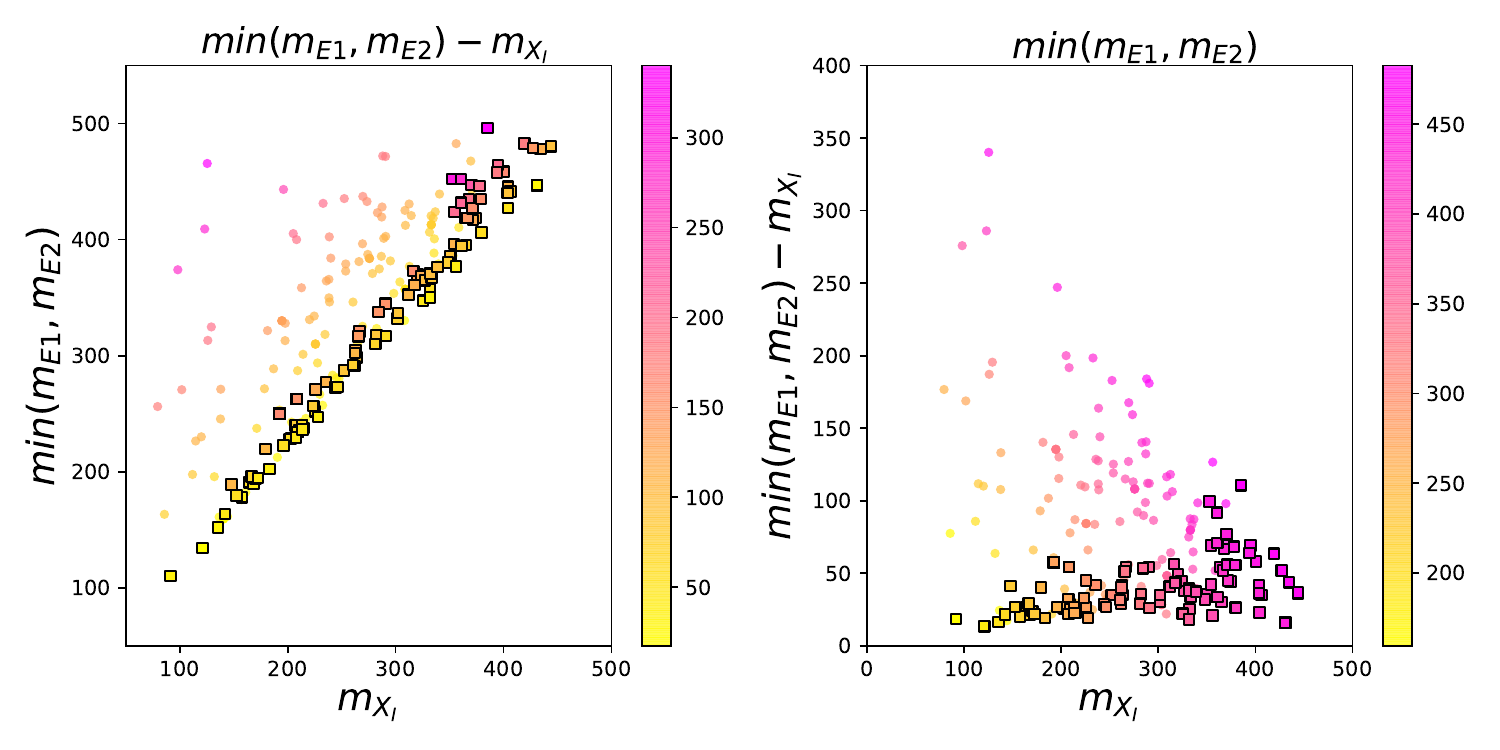,height=8.cm}
 %\end{center}
\vspace{-0.5cm} \caption{All the samples achieve FOFP and accommodate the DM relic
density, muon $g-2$ anomaly, and various constraints mentioned previously. The circles and squares are excluded and allowed by the direct searches for $2\ell+E^{miss}_T$ at the LHC. 
}
 \label{figlhc}
\end{figure}
%%%%%%%%%%%%%%%%%%%%

In Fig.  \ref{figlhc}, we employ the ATLAS analysis of $2\ell+E_T^{miss}$ at the LHC to restrict the parameter space which has been satisfied by the DM relic density,
muon $g-2$, FOPT, and various constraints discussed above.  Fig. \ref{figlhc} indicates that the DM mass is allowed 
to be as low as 100 GeV when the value of $min(m_{E_1},~m_{E_2})-m_{X_I}$ is small. 
This is because the muon from the vector-like lepton decay has soft energy, and its detection efficiency is decreased at the LHC.
As the DM mass increase, the value of $min(m_{E_1},~m_{E_2})-m_{X_I}$ increases, and the
lightest charged vector-like lepton is allowed to have a more large mass.

\subsection{Gravitational wave signature}
%%%% 1812.08293
Stochastic GWs are produced during a FOPT via bubble collision, sound waves
in the plasma  and the magneto-hydrodynamics turbulence. Since most of the PT energy
is pumped into the surrounding fluid shells, making sound waves the dominant contribution
to GWs, we will focus on the GW spectrum from the sound waves in the plasma.

The sound wave spectra can be expressed as functions of two FOPT
parameters $\beta$ and $\alpha$, 
\begin{eqnarray}
\frac{\beta}{H_n}=T\frac{d (S_3(T)/T)}{d T}|_{T=T_n}\; ,~~~~ \alpha=\frac{\Delta\rho}{\rho_R}=\frac{\Delta\rho}{\pi^2 g_{\ast} T_n^4/30}\;.
\end{eqnarray}
Where $H_n$ is the Hubble parameter at the nucleation temperature $T_n$, and $g_{\ast}$ is the effective number of relativistic degrees of freedom.
$\beta$ characterizes roughly the inverse time duration of the strong first-order PT, and 
$\alpha$ is defined as the vacuum energy released from the PT normalized by the total radiation energy
density $\rho_R$ at $T_n$.

The GW spectrum from the sound waves can be expressed by \cite{gw-sw}
\begin{eqnarray}
\Omega_{\textrm{sw}}h^{2} & \ = \ &
2.65\times10^{-6}\left( \frac{H_{n}}{\beta}\right)\left(\frac{\kappa_{v} \alpha}{1+\alpha} \right)^{2}
\left( \frac{100}{g_{\ast}}\right)^{1/3} v_w\nonumber \\
&&\times  \left(\frac{f}{f_{sw}} \right)^{3} \left( \frac{7}{4+3(f/f_{\textrm{sw}})^{2}} \right) ^{7/2} \Upsilon(\tau_{sw})\ ,
\label{eq:soundwaves}
\end{eqnarray}
where $v_w$ is the wall velocity, and we take $v_w=c_s=\sqrt{1/3}$ with $c_s$ being the sound velocity.
 $f_{\text{sw}}$ is the present peak frequency of the spectrum,
\begin{equation}
f_{\textrm{sw}} \ = \ 
1.9\times10^{-5}\frac{1}{v_w}\left(\frac{\beta}{H_{n}} \right) \left( \frac{T_{n}}{100\textrm{GeV}} \right) \left( \frac{g_{\ast}}{100}\right)^{1/6} \textrm{Hz} \,.
\label{fsw}
\end{equation}
The $\kappa_{v}$ is the fraction of latent heat transformed into the kinetic energy of the fluid \cite{1004.4187},
\beq
\kappa_v \simeq  \frac{\alpha^{2/5}}
{0.017 + (0.997 + \alpha)^{2/5}}.
\eeq
The suppression factor of Eq. (\ref{eq:soundwaves}) \cite{2007.08537}
\beq
\Upsilon(\tau_{sw})=1-\frac{1}{\sqrt{1+2\tau_{sw}H_n}},
\eeq 
appears due to the finite lifetime $\tau_{sw}$ of the sound waves \cite{2003.07360,2003.08892},
\beq
\tau_{sw}=\frac{\tilde{v}_W(8\pi)^{1/3}}{\beta \bar{U}_f}, ~~ \bar{U}^2_f=\frac{3}{4}\frac{\kappa_v\alpha}{1+\alpha}.
\eeq

%%%%%%%%%%%%%%%%%%%%%
\begin{figure}
 \epsfig{file=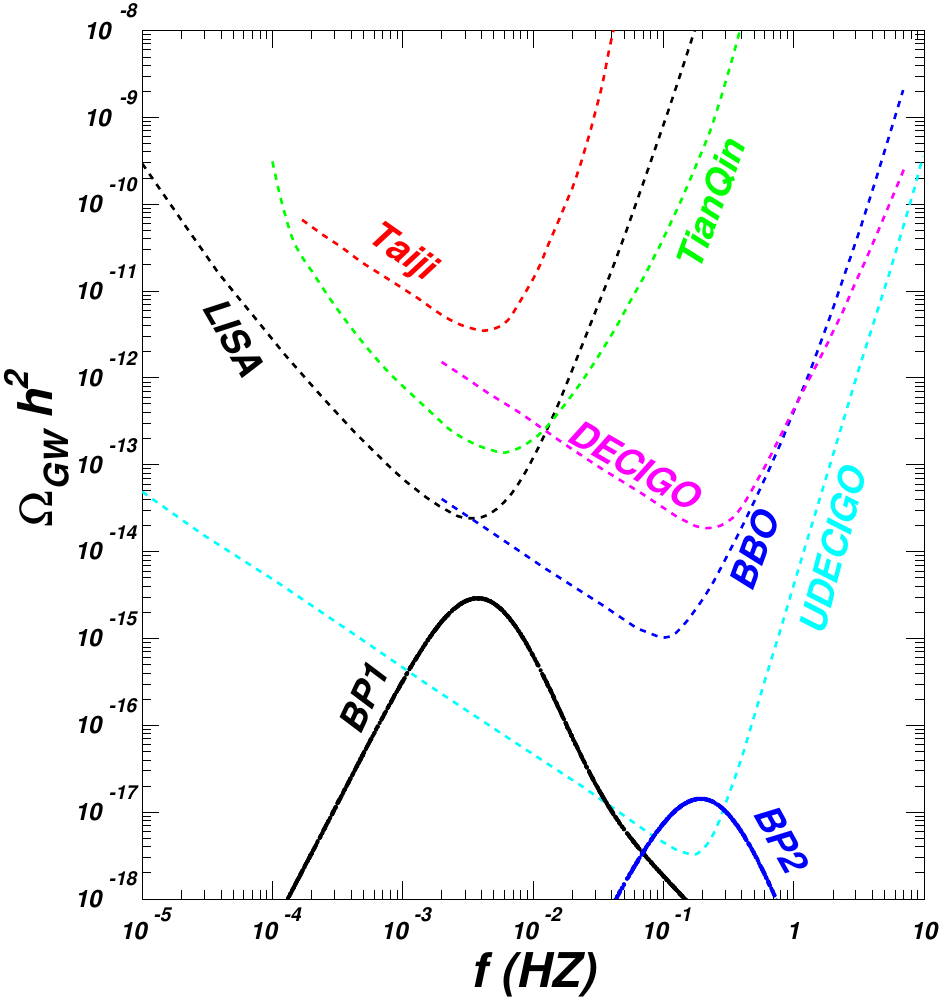,height=8.cm}
 %\end{center}
\vspace{-0.5cm} \caption{Gravitational wave spectra for the BP1 and BP2. 
}
 \label{figgw}
\end{figure}
%%%%%%%%%%%%%%%%%%%%
We calculate GW spectra for thousands of parameter points accommodating the muon $g-2$, DM relic density, and the exclusion limits of the LHC direct searches, and
find that all the peak strengths are below the sensitivity curve of BBO.
About 10 percent of the survived points can generate U-DECIGO sensitive gravitational wave, including BP1 and BP2.
These points favor a small $\lambda_S$ for which the strength of FOPT tends to have a large value.
The GW spectra of BP1 and BP2 are shown along with expected sensitivities of various future
interferometer experiments in Fig. \ref{figgw}.
The lowest peak frequency is 0.003 HZ from BP1, and the highest peak frequency is 0.2 HZ from BP2.

\section{Conclusion}
 We study an extra $U(1)_{L_\mu-L_\tau}$ gauge symmetry extension of the standard model by considering the dark matter, muon $g-2$ anomaly, the $U(1)_{L_\mu-L_\tau}$ breaking PT, 
GW spectra, and the bound from the direct detection at the LHC.
We obtained the following observations:
(i)
A joint explanation of the dark matter relic density and muon $g-2$ anomaly rules out the region where both $min(m_{E_1},m_{E_2},m_N,m_{X_R})$ and $min(m_{Z'},m_S)$ 
are much larger than $m_{X_I}$. (ii) A first-order $U(1)_{L_\mu-L_\tau}$ breaking PT can be achieved in the parameter space explaining
the DM relic density and muon $g-2$ anomaly simultaneously, and the corresponding gravitational wave spectra can reach the sensitivity of U-DECIGO.
(iii) The mass spectra of the vector-like leptons and dark matter are stringently restricted by the direct searches at the LHC.

\section*{Acknowledgment}
This work was supported by the National Natural Science Foundation
of China under grant 11975013.

\end{document}